\documentclass[prl,superscriptaddress,10pt]{revtex4}%
\usepackage[ansinew]{inputenc}
\usepackage{graphicx}
\usepackage{amsmath}
\usepackage{amsthm}
\usepackage{bm}
\usepackage{layout}
\usepackage{float}
\usepackage{amsfonts}
\usepackage{amssymb}%
\normalsize
\begin{document}

\title{Logical independence and quantum randomness}

\begin{abstract}
We propose a link between logical independence and quantum physics. We
demonstrate that quantum systems in the eigenstates of Pauli group operators
are capable of encoding mathematical axioms and show that Pauli group quantum
measurements are capable of revealing whether or not a given proposition is
logically dependent on the axiomatic system. Whenever a mathematical
proposition is logically independent of the axioms encoded in the measured
state, the measurement associated with the proposition gives random outcomes.
This allows for an experimental test of logical independence. Conversely, it
also allows for an explanation of the probabilities of random outcomes
observed in Pauli group measurements from logical independence without
invoking quantum theory. The axiomatic systems we study can be completed and
are therefore \textit{not} subject to Gödel's incompleteness theorem.

\end{abstract}
\date{\today}%

\author{Tomasz Paterek}%

\altaffiliation{Present address: \textit{Centre for Quantum Technologies, National University of Singapore, 3 Science Drive 2, 117543 Singapore, Singapore}}%

\affiliation
{Institut f\"ur Quantenoptik und Quanteninformation (IQOQI), \"Osterreichische Akademie der Wissenschaften,\\ Boltzmanngasse 3, 1090 Wien, Austria}%
%

\author{Johannes Kofler}%
%

\affiliation
{Institut f\"ur Quantenoptik und Quanteninformation (IQOQI), \"Osterreichische Akademie der Wissenschaften,\\ Boltzmanngasse 3, 1090 Wien, Austria}%
%

\affiliation{Fakult\"at f\"ur Physik, Universit\"{a}%
t Wien, Boltzmanngasse 5, 1090 Wien, Austria}%
\author{Robert Prevedel}%
%

\affiliation{Fakult\"at f\"ur Physik, Universit\"{a}%
t Wien, Boltzmanngasse 5, 1090 Wien, Austria}%
%

\author{Peter Klimek}%

\altaffiliation{Present address: \textit{Complex Systems Research
Group, HNO, Medical University of Vienna, W\"ahringer G\"urtel
18-20, 1090 Wien, Austria}}%

\affiliation{Fakult\"at f\"ur Physik, Universit\"{a}%
t Wien, Boltzmanngasse 5, 1090 Wien, Austria}%
%

\author{Markus Aspelmeyer}%
%

\affiliation
{Institut f\"ur Quantenoptik und Quanteninformation (IQOQI), \"Osterreichische Akademie der Wissenschaften,\\ Boltzmanngasse 3, 1090 Wien, Austria}%
%

\affiliation{Fakult\"at f\"ur Physik, Universit\"{a}%
t Wien, Boltzmanngasse 5, 1090 Wien, Austria}%
%

\author{Anton Zeilinger}%
%

\affiliation
{Institut f\"ur Quantenoptik und Quanteninformation (IQOQI), \"Osterreichische Akademie der Wissenschaften,\\ Boltzmanngasse 3, 1090 Wien, Austria}%
%

\affiliation{Fakult\"at f\"ur Physik, Universit\"{a}%
t Wien, Boltzmanngasse 5, 1090 Wien, Austria}%
%

\author{{\v C}aslav Brukner}%
%

\affiliation
{Institut f\"ur Quantenoptik und Quanteninformation (IQOQI), \"Osterreichische Akademie der Wissenschaften,\\ Boltzmanngasse 3, 1090 Wien, Austria}%
%

\affiliation{Fakult\"at f\"ur Physik, Universit\"{a}%
t Wien, Boltzmanngasse 5, 1090 Wien, Austria}%
%

\maketitle
As opposed to the case of classical statistical physics, the theorems by
Kochen and Specker \cite{Koch1967} and Bell \cite{Bell1964} opened up the
possibility to view probabilities in quantum physics as irreducible and not as
stemming from our ignorance about some (non-contextual or local)
predeterminated properties. Adopting this view, one can ask if there is any
reason why such irreducible probabilities should have different values at all.
Here we show that---at least in a certain subset of measurements (Pauli group
measurements)---quantum probabilities can be seen as following from logical
independence of mathematical propositions which are associated to the
measurements without invoking quantum theory itself.

Any formal system is based on axioms, which are propositions that are defined
to be true. Whenever a proposition and a given set of axioms together contain
more information than the axioms themselves, the proposition can neither be
proved nor disproved from the axioms -- it is \textit{logically independent}
(or mathematically undecidable \cite{Chai1982,Calu2005}). If a proposition is
independent of the axioms, neither the proposition itself nor its negation
creates an inconsistency together with the axiomatic system.

We demonstrate that the states of quantum systems are capable of encoding
mathematical axioms. Quantum mechanics imposes an upper limit on how much
information can be encoded in a quantum state \cite{Hole1973,Zeil1999}, thus
limiting the information content of the set of axioms. We show that quantum
measurements are capable of revealing whether a given proposition is
independent or not of the set of axioms. Whenever a mathematical proposition
is independent of the axioms encoded in the state, the measurement associated
with the proposition gives random outcomes. This allows for an
\textit{experimental} test of logical independence by realizing in the
laboratory both the actual quantum states and the required quantum
measurements. Our axiomatic systems can be completed and are therefore
\textit{not} subject to Gödel's incompleteness theorem
\cite{Goed1931,Nage1960}.

Intuitively, independent propositions contain entirely \textit{new
information} which cannot be reduced to the information in the axioms. This
point of view is related to Chaitin's information-theoretical formulation of
logical independence \cite{Chai1982,Calu2005}: Given a set of axioms that
contains a certain amount of information, it is impossible to deduce the truth
value of a proposition which, together with the axioms, contains more
information than the set of axioms itself.

To give an example, consider Boolean functions of a single binary argument:%
\begin{equation}
x\in\{0,1\}\;\rightarrow\;y=f(x)\in\{0,1\}
\end{equation}
There are four such functions, $y_{k}$ ($k=0,1,2,3$), shown in
Figure~\ref{Figure_Undecidability_functions}. We shall discuss the following
(binary) propositions about their properties:\begin{figure}[t]
\begin{center}
\includegraphics[width=.225\textwidth]{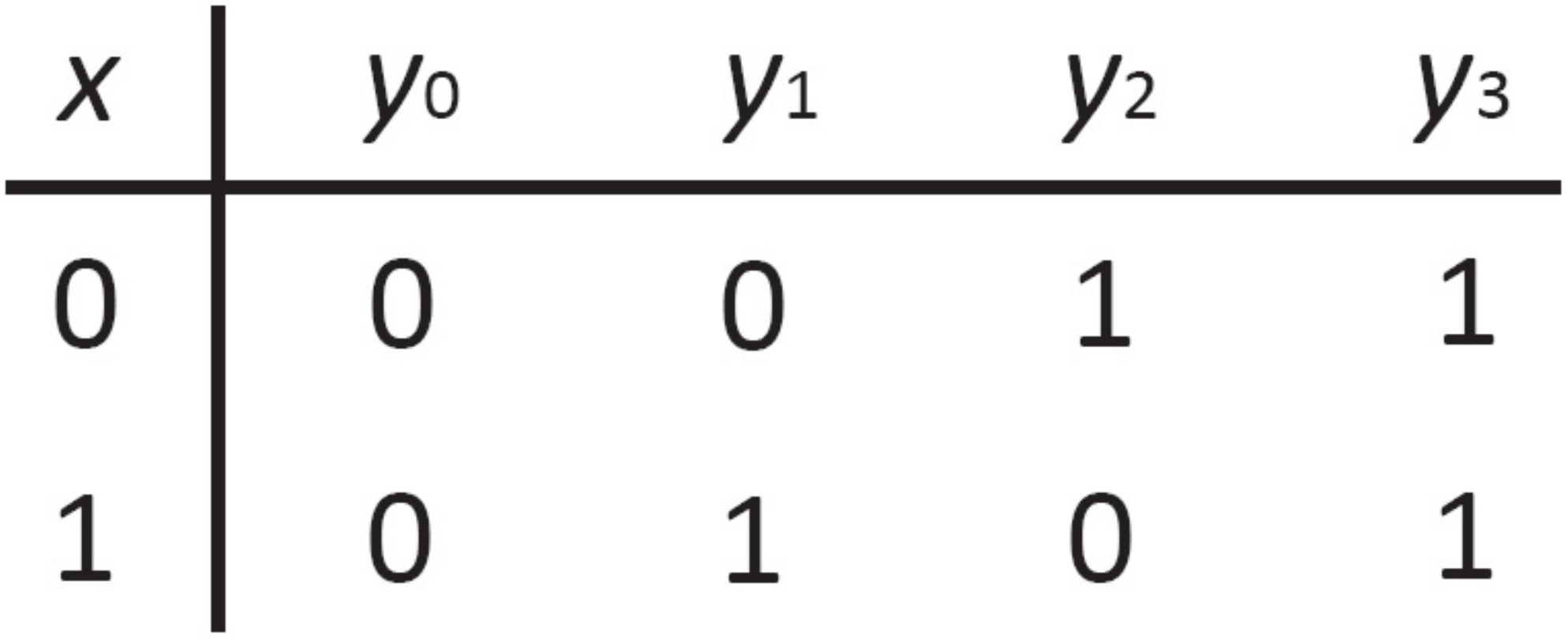}
\end{center}
\par
\vspace{-0.25cm}\caption{The four Boolean functions $y=f(x)$ of a binary
argument, i.e.\ $f(x)=0,1$ with $x=0,1$. The different functions are labeled
by $y_{k}$ with $k=0,1,2,3$.}%
\label{Figure_Undecidability_functions}%
\end{figure}%
\begin{align*}
\text{(A)}\quad``\text{The value of }f(0)\text{ is `0', i.e.~}f(0)  &
=0\text{.}\textquotedblright\\
\text{(B)}\quad``\text{The value of }f(1)\text{ is `0', i.e.~}f(1)  &
=0\text{.}\textquotedblright%
\end{align*}
These two propositions are logically independent. Knowing the truth value of
one of them does not allow to infer the truth value of the other. Ascribing
truth values to both propositions requires two bits of information. If one
postulates only proposition (A) to be true, i.e. if one chooses (A) as an
`axiom', then it is impossible to prove proposition (B) from (A). Having only
axiom (A), i.e. only this one bit of information, there is \textit{not enough
information} to know also the truth value of (B). Hence, proposition (B) is
logically independent from the system containing the single axiom (A). Another
example of an independent proposition within the same axiomatic system is:%
\[
\text{(C)}\quad``\text{The function is constant, i.e.~}f(0)=f(1)\text{.}%
\textquotedblright%
\]
Again, this statement can neither be proved nor disproved from the axiom (A)
alone because (C) is independent of (A) as it involves $f(1)$.

We refer to such independent propositions to which one cannot simultaneously
ascribe definite truth values -- given a limited amount of information
resources -- as \textit{logically complementary propositions}. Knowing the
truth value of one of them precludes any knowledge about the others. Given the
limitation of one bit of information encoded in the axiom, all three
propositions (A), (B) and (C) are logically complementary to each other.

When the information content of the axioms and the number of independent
propositions increase, more possibilities arise. Already the case of two bits
as the information content is instructive. Consider two independent Boolean
functions $f_{1}(x)$ and $f_{2}(x)$ of a binary argument. The two bits may be
used to define properties of the individual functions or they may define joint
features of the functions. An example of the first type is the following
two-bit proposition:%
\begin{align*}
\text{(D)}\quad``\text{The value of }f_{1}(0)\text{ is `0', i.e.~}f_{1}(0)  &
=0\text{.}\textquotedblright\\
``\text{The value of }f_{2}(1)\text{ is `0', i.e.~}f_{2}(1)  &  =0\text{.}%
\textquotedblright%
\end{align*}
An example of the second type is:%
\begin{align*}
\text{(E)}\quad``\text{The functions have the same value for argument `0',
i.e.~}f_{1}(0)  &  =f_{2}(0)\text{.}\textquotedblright\\
``\text{The functions have the same value for argument `1', i.e.~}f_{1}(1)  &
=f_{2}(1)\text{.}\textquotedblright%
\end{align*}
Both (D) and (E) consist of two elementary (binary) propositions. Their truth
values are of the form of vectors with two components being the truth values
of their elementary propositions. The propositions (D) and (E) are logically
complementary. Given (E) as a two-bit axiom, all the \textit{individual}
function values remain undefined and thus one can determine neither of the two
truth values of elementary propositions in (D).

A qualitatively new aspect of multi-bit axioms is the existence of
\textquotedblleft partially\textquotedblright\ independent propositions,
i.e.\ propositions that contain more than one elementary proposition of which
only some are independent. An example of such a partially independent
proposition within the system consisting of the two-bit axiom (D) is:%
\begin{align*}
\text{(F)}\quad``\text{The value of }f_{1}(0)\text{ is `0', i.e.~}f_{1}(0)  &
=0\text{.}\textquotedblright\\
``\text{The value of }f_{2}(0)\text{ is `0', i.e.~}f_{2}(0)  &  =0\text{.}%
\textquotedblright%
\end{align*}
The first elementary proposition is the same as in (D) and thus it is
definitely true. The impossibility to decide the second elementary proposition
leads to partial independence of proposition (F). In a similar way,
proposition (F) is partially independent of the axiomatic system of (E).

The discussion so far was purely \textit{mathematical}. We have described
finite axiomatic systems (of limited information content) using properties of
Boolean functions. Now we show that the independence of mathematical
propositions can be tested in quantum experiments. To this end we introduce a
\textit{physical} \textquotedblleft black box\textquotedblright\ whose
internal configuration encodes Boolean functions. The black box hence forms a
bridge between mathematics and physics. Quantum systems enter it and the
properties of the functions, i.e.\ the truth values of propositions, are
written onto the quantum states of the systems. Finally, measurements
performed on the systems extract information about the properties of the
configuration of the black box and thus about the properties of the functions.

We begin with the simplest case of a qubit (e.g. a spin-%
${\frac12}$
particle or the polarization of a photon) entering the black box in a
well-defined state and a single bit-to-bit function $f(x)$ encoded in the
black box. Inside the black box two subsequent operations alter the state of
the input qubit. The first operation encodes the value of $f(1)$ via
application of $\hat{\sigma}_{z}^{f(1)}$, i.e.\ the Pauli $z$-operator taken
to the power of $f(1)$. The second operation encodes $f(0)$ with $\hat{\sigma
}_{x}^{f(0)}$, i.e.\ the Pauli $x$-operator taken to the power of $f(0)$. The
total action of the black box is%
\begin{equation}
\hat{U}=\hat{\sigma}_{x}^{f(0)}\,\hat{\sigma}_{z}^{f(1)}\,. \label{eq U}%
\end{equation}

Consider the input qubit to be in one of the eigenstates of the Pauli operator
i$^{mn}\,\hat{\sigma}_{x}^{m}\,\hat{\sigma}_{z}^{n}$ (with i the imaginary
unit). The three particular choices $(m,n)=(0,1)$, $(1,0)$, or $(1,1)$
correspond to the three Pauli operators along orthogonal directions (in the
Bloch sphere) $\hat{\sigma}_{z}$, $\hat{\sigma}_{x}$, or $\hat{\sigma}_{y}%
=\;$i$\,\hat{\sigma}_{x}\,\hat{\sigma}_{z}$, respectively. The measurements of
these operators are quantum complementary: Given a system in an eigenstate of
one of them, the results of the other measurements are totally random. The
input density matrix reads%
\begin{equation}
\hat{\rho}=\tfrac{1}{2}%
\,[\leavevmode\hbox{\small1\kern-3.3pt\normalsize1}+\lambda_{mn}%
\,\text{i}^{mn}\,\hat{\sigma}_{x}^{m}\,\hat{\sigma}_{z}^{n}]\,,
\end{equation}
with $\lambda_{mn}=\pm1$ and
$\leavevmode\hbox{\small1\kern-3.3pt\normalsize1}$ the identity operator. It
evolves under the action of the black box to%
\begin{equation}
\hat{U}\hat{\rho}\,\hat{U}^{\dag}=\tfrac{1}{2}%
\,[\leavevmode\hbox{\small1\kern-3.3pt\normalsize1}+\lambda_{mn}%
\,(-1)^{nf(0)+mf(1)}\,\text{i}^{mn}\,\hat{\sigma}_{x}^{m}\,\hat{\sigma}%
_{z}^{n}]\,.
\end{equation}

Depending on the value of $n\,f(0)+m\,f(1)$ (throughout the paper all sums are
taken modulo 2), the state after the black box is either the same or
orthogonal to the initial one. If one now performs a measurement in the basis
of the initial state, (i.e.\ the eigenbasis of the operator i$^{mn}%
\,\hat{\sigma}_{x}^{m}\,\hat{\sigma}_{z}^{n}$), the outcome reveals the value
of $n\,f(0)+m\,f(1)$ and hence the measurement can be considered as
\textit{checking the truth value of the proposition}%
\[
\text{(G)}\quad``n\,f(0)+m\,f(1)=0\text{.}\textquotedblright%
\]

It is crucial to note that each of the three quantum complementary
measurements $\hat{\sigma}_{z}$, $\hat{\sigma}_{x}$, or $\hat{\sigma}_{y}$ --
given the suitable initial state -- reveals the truth value of one of the
independent propositions (A), (B), or (C), respectively.

\textit{Independent} of the initial state, we now identify the quantum
measurement $(m,n)$ with the question about the truth value of the
corresponding mathematical proposition (G). Those states that give a definite
(i.e.\ not random) outcome in the quantum measurement encode (G) or its
negation as an axiom. For example, the two eigenstates of $\hat{\sigma}_{z}$
after the black box encode (A) or its negation as an axiom, and the
$\hat{\sigma}_{z}$ measurement reveals the truth value of the proposition (A).
This one bit is the maximal amount of information that can be encoded in a
qubit \cite{Hole1973,Zeil1999}.

\begin{figure}[t]
\begin{center}
\includegraphics{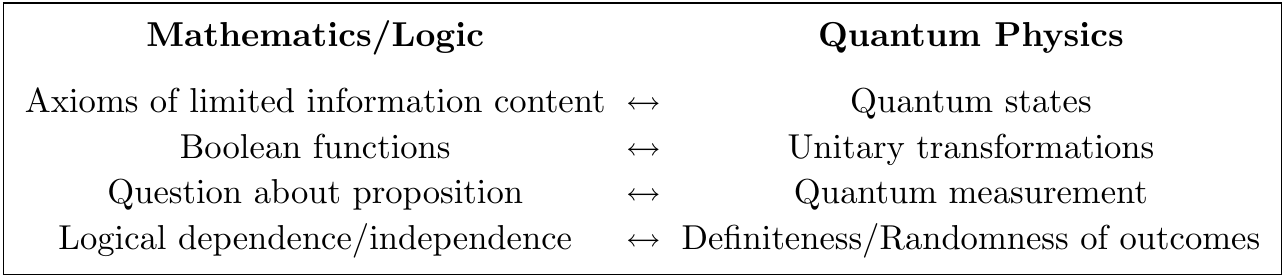}
\end{center}
\par
\vspace{-0.25cm}\caption{The link between logical independence and quantum
randomness.}%
\label{Figure_Undecidability_table}%
\end{figure}

When a physical system prepared in an eigenstate of a Pauli operator is
measured along complementary directions, the measurement outcomes are
\textit{random}. Correspondingly, the proposition identified with a
complementary observable is \textit{independent} from the one-bit axiom
encoded in the measured state. For example, the measurement of $\hat{\sigma
}_{x}$ on an eigenstate of $\hat{\sigma}_{z}$ gives random outcomes, and
accordingly proposition (B) is independent of the one-bit axiom (A). This
links logical independence and quantum randomness in complementary
measurements. We propose that it is therefore possible to
\textit{experimentally} find out whether a proposition is logically
independent or not, as summarized in Figure~\ref{Figure_Undecidability_table}%
.

In a single experimental run it is impossible to infer whether the outcome is
definite or random and thus whether it stemmed from a dependent or independent
proposition. Therefore, any quantum experiment revealing logical independence
requires many runs. (It can be shown that -- given a certain level of noise --
the probability to infer wrongly whether the proposition is dependent or not
decays exponentially with the length of the outcome string.)

Generalizing the above reasoning to multiple qubits, we show in the following
that \textit{whenever} the proposition identified with a Pauli group
measurement is dependent (on the axioms encoded into the qubits), the
measurement outcome is definite, and whenever it is independent, the
measurement outcome is random. Consider $N$ black boxes, one for each qubit.
They encode $N$ Boolean functions $f_{j}(x)$ numbered by $j=1,\ldots,N$ by
applying the operation%
\begin{equation}
\hat{U}_{N}=\hat{\sigma}_{x}^{f_{1}(0)}\,\hat{\sigma}_{z}^{f_{1}(1)}%
\otimes\cdots\otimes\hat{\sigma}_{x}^{f_{N}(0)}\,\hat{\sigma}_{z}^{f_{N}%
(1)}\,.
\end{equation}
The initial $N$-qubit state is chosen to be a particular one of the $2^{N}$
eigenstates of certain $N$ \textit{independent and mutually commuting} tensor
products of Pauli operators, numbered by $p=1,...,N$:%
\begin{equation}
\hat{\Omega}_{p}\equiv\text{i}^{m_{1}(p)n_{1}(p)}\,\hat{\sigma}_{x}^{m_{1}%
(p)}\,\hat{\sigma}_{z}^{n_{1}(p)}\otimes\cdots\otimes\text{i}^{m_{N}%
(p)n_{N}(p)}\,\hat{\sigma}_{x}^{m_{N}(p)}\,\hat{\sigma}_{z}^{n_{N}(p)}\,,
\end{equation}
with $m_{j}(p),n_{j}(p)\in\{0,1\}$. A broad family of such states is the
family of stabilizer \cite{Gott1996,Niel2000} and graph states \cite{Raus2003}%
. (Note that not all states can be described within this framework.) As
before, each qubit propagates through its black box. After leaving them, the
qubits' state encodes the truth values of the following $N$ independent binary
propositions (negating the false propositions, one has $N$ true ones which
serve as axioms):%
\[
\text{(H}_{p}\text{)}\quad``%
{\textstyle\sum\nolimits_{j=1}^{N}}
[n_{j}(p)\,f_{j}(0)+m_{j}(p)\,f_{j}(1)]=0\text{.}\textquotedblright%
\]
In suitable measurements quantum mechanics provides a way to test whether
certain propositions are dependent or not. If one measures the operator of the
Pauli group \cite{Niel2000}%
\begin{equation}
\hat{\Theta}\equiv\text{i}^{\alpha_{1}\beta_{1}}\,\hat{\sigma}_{x}^{\alpha
_{1}}\,\hat{\sigma}_{z}^{\beta_{1}}\otimes\cdots\otimes\text{i}^{\alpha
_{N}\beta_{N}}\,\hat{\sigma}_{x}^{\alpha_{N}}\,\hat{\sigma}_{z}^{\beta_{N}}\,,
\end{equation}
with $\alpha_{j},\beta_{j}\in\{0,1\}$, one tests whether the proposition%
\[
\text{(J)}\quad``%
{\textstyle\sum\nolimits_{j=1}^{N}}
[\beta_{j}\,f_{j}(0)+\alpha_{j}\,f_{j}(1)]=0\text{.}\textquotedblright%
\]
is dependent or not. The proposition (J) can be represented as the
$2N$-dimensional proposition vector $\vec{J}=(\alpha_{1},\ldots,\alpha
_{N},\beta_{1},\ldots,\beta_{N})$ with binary entries. Therefore, there are
$4^{N}$ different (J)'s. For all dependent propositions, the vectors $\vec{J}$
are linear combinations of the vectors $\vec{H}_{p}=(m_{1}(p),\ldots
,m_{N}(p),n_{1}(p),\ldots,n_{N}(p))$ representing the axioms, i.e.\ $\vec{J}=%
{\textstyle\sum\nolimits_{p=1}^{N}}
k_{p}\vec{H}_{p}$. Since $\alpha_{j},\beta_{j}$ are binary, the coefficients
must also be binary: $k_{p}\in\{0,1\}$. This gives $2^{N}$ dependent
propositions (J). The corresponding operators $\hat{\Theta}$ can be written as
the products $\hat{\Omega}_{1}^{k_{1}}\cdots\hat{\Omega}_{N}^{k_{N}}$. In this
case $\hat{\Theta}$ commutes with all the $\hat{\Omega}_{p}$'s, and the
quantum mechanical formalism implies that the measurement of $\hat{\Theta}$
has a definite outcome. The measurements of all the remaining $4^{N}%
-2^{N}=2^{N}(2^{N}-1)$ operators $\hat{\Theta}$ give random outcomes, and the
corresponding propositions (J) are independent. Note that there are many more
independent propositions of the form (J) than dependent ones. The ratio
between their numbers increases exponentially with the number of qubits,
i.e.\ $\frac{2^{N}(2^{N}-1)}{2^{N}}=O(2^{N})$.

In logic, one can always complete the axiomatic system by adding new axioms to
the set of (H$_{p}$) such that any proposition (J) becomes dependent. However,
this would require the axioms to be encoded in more than $N$ qubits. Having
only $N$ qubits, projecting these qubits into new quantum states, and
propagating them through their black boxes, new propositions can become axioms
but only if some or all previous axioms become independent propositions. This
is a consequence of the limited information content of the quantum system.

We have proved that a proposition of the type (J) is dependent on the
axiomatic system (H$_{p}$) if and only if the corresponding measurement
$\hat{\Theta}$ from the Pauli group is commuting with all $\hat{\Omega}_{p}$.
Note that one does not need to first prove the (in)dependence of a proposition
by logic before one is able to identify the experiment to test it. For a given
set of (H$_{p}$), defining an $N$-bit axiom, one must prepare a joint
eigenstate of $N$ commuting operators $\hat{\Omega}_{p}$. In order to test the
logical (in)dependence of a new proposition (J), one needs to measure the
operator $\hat{\Theta}$ that corresponds to (J) in this state. The procedures
of preparation and measurement can be performed without knowing whether (J) is
logically independent of the set of (H$_{p}$).

The measurement $\hat{\Theta}$ is highly degenerate because it tests the
logical (in)dependence of the binary proposition (J) of the axioms. Less
degenerate measurements are possible which simultaneously test the logical
(in)dependence of several elementary propositions of the form (J). Such
multi-bit propositions contain many elementary propositions. If not all of
them are independent of the axioms, this gives rise to partial independence.
This provides an explanation for different values of outcome probabilities in
Pauli group measurements, which is based on logic without invoking quantum
theory. A measurement corresponding to any single independent elementary
proposition (with two possible measurement outcomes) gives uniformly random
results. In a measurement whose outcomes reveal the independence of $m$
independent elementary propositions (with $2^{m}$ possible measurement
outcomes) these outcomes occur with probabilities $\frac{1}{2^{m}}$. (The
results revealing dependence of elementary propositions are always definite.)

To illustrate the idea of multi-bit propositions and partial independence,
consider the two bits of proposition (E) described above correspond to the set
of independent commuting operators $\hat{\Omega}_{1}=\hat{\sigma}_{z}%
\otimes\hat{\sigma}_{z}$ and $\hat{\Omega}_{2}=\hat{\sigma}_{x}\otimes
\hat{\sigma}_{x}$. The common eigenbasis of these operators is spanned by the
maximally entangled Bell states (basis $b_{\text{E}}$): $\left\vert \Phi^{\pm
}\right\rangle =\tfrac{1}{\sqrt{2}}\left(  \left\vert z+\right\rangle
_{1}\left\vert z+\right\rangle _{2}\pm\left\vert z-\right\rangle
_{1}\left\vert z-\right\rangle _{2}\right)  $, $\left\vert \Psi^{\pm
}\right\rangle =\tfrac{1}{\sqrt{2}}\left(  \left\vert z+\right\rangle
_{1}\left\vert z-\right\rangle _{2}\pm\left\vert z-\right\rangle
_{1}\left\vert z+\right\rangle _{2}\right)  $, where e.g.\ $\left\vert
z\pm\right\rangle _{1}$ denotes the eigenstate with the eigenvalue $\pm1$ of
$\hat{\sigma}_{z}$ for the first qubit. Thus, after the black boxes the four
Bell states encode the four possible truth values of the elementary
propositions in (E) and a so-called Bell State Analyzer \cite{Wein1994}
(i.e.\ an apparatus that measures in the Bell basis) reveals these values. In
the same way, the truth values of the elementary propositions in (F) are
encoded in the eigenstates of local $\hat{\sigma}_{z}$ bases, i.e.\ by the
four states $\left\vert z\pm\right\rangle _{1}\left\vert z\pm\right\rangle
_{2}$ (basis $b_{\text{F}}$). Finally, the elementary propositions in (D) are
linked with the four product states $\left\vert z\pm\right\rangle
_{1}\left\vert x\pm\right\rangle _{2}$ (basis $b_{\text{D}}$). In general, if
all the axioms involve joint properties of Boolean functions the multi-partite
state encoding these axioms must be entangled.

Measurements in the Bell basis, $b_{\text{E}}$, prove that the entangled state
indeed encodes joint properties of the two functions, i.e.\ information about
(E). Measurements in other bases can then be interpreted in terms of
\textquotedblleft partial\textquotedblright\ and \textquotedblleft
full\textquotedblright\ independence. Proposition (D) is fully independent
given (E) as an axiom and the four possible measurement results are completely
random with probabilities of $\tfrac{1}{4}$. On the other hand, proposition
(F) is partially independent, which is disclosed by the fact that two (out of
four) outcomes never occur, while the two remaining occur randomly, i.e.\ each
with probability $\tfrac{1}{2}$.

When the outcome of a quantum measurement is definite, it need not possess an
\textit{a priori} relation to the actual truth value of a dependent
proposition as imposed by classical logic. This can be demonstrated for three
qubits initially in the Greenberger-Horne-Zeilinger (GHZ) state
\cite{Gree1989}%
\begin{equation}
\left\vert \text{GHZ}\right\rangle =\tfrac{1}{\sqrt{2}}\left(  \left\vert
z+\right\rangle _{1}\left\vert z+\right\rangle _{2}\left\vert z+\right\rangle
_{3}+\left\vert z-\right\rangle _{1}\left\vert z-\right\rangle _{2}\left\vert
z-\right\rangle _{3}\right)  .
\end{equation}
We choose as axioms the propositions%
\begin{align*}
\text{(K}_{1}\text{)}\quad``f_{1}(0)+f_{1}(1)+f_{2}(0)+f_{2}(1)\quad\quad
\quad\;\,+f_{3}(1)  &  =1\text{.}\textquotedblright\\
\text{(K}_{2}\text{)}\quad``f_{1}(0)+f_{1}(1)\quad\quad\quad\;\,+f_{2}%
(1)+f_{3}(0)+f_{3}(1)  &  =1\text{.}\textquotedblright\\
\text{(K}_{3}\text{)}\quad``\quad\quad\quad\;\,f_{1}(1)+f_{2}(0)+f_{2}%
(1)+f_{3}(0)+f_{3}(1)  &  =1\text{.}\textquotedblright%
\end{align*}
linked with the operators $\hat{\sigma}_{y}\otimes\hat{\sigma}_{y}\otimes
\hat{\sigma}_{x}$, $\hat{\sigma}_{y}\otimes\hat{\sigma}_{x}\otimes\hat{\sigma
}_{y}$, and $\hat{\sigma}_{x}\otimes\hat{\sigma}_{y}\otimes\hat{\sigma}_{y}$,
respectively. One can \textit{logically} derive from (K$_{1}$) to (K$_{3}$)
the true proposition%
\[
\text{(L)}\quad``f_{1}(1)+f_{2}(1)+f_{3}(1)=1\text{.}\textquotedblright%
\]
On the other hand, the proposition (L) is identified with the measurement of
$\hat{\sigma}_{x}\otimes\hat{\sigma}_{x}\otimes\hat{\sigma}_{x}$, but the
result imposed by quantum mechanics corresponds to the \textit{negation} of
(L), namely: \textquotedblleft$f_{1}(1)+f_{2}(1)+f_{3}(1)=0$%
.\textquotedblright\ This is the heart of the GHZ
argument~\cite{Gree1989,Merm1990,Pan2000}. In the (standard logical)
derivation of (L) the individual function values are well defined and the
\textit{same}, independently of the axiom (K$_{i}$) in which they appear.
Since this is equivalent to the assumption of non-contextuality
\cite{Koch1967,Pere1995}, the truth values of dependent propositions found in
quantum experiments do not necessarily have to be the same as the ones derived
by classical logic. Nonetheless, there is a one-to-one correspondence between
definiteness (or randomness) of the measurement outcomes and the associated
propositions being dependent (or independent) within axiomatic set. As shown
above, this correspondence is independent of the rules used to infer the
specific truth values of the propositions (e.g.\ classical logic or quantum theory).

One might raise the question whether a classical device can be constructed to
reveal the independence of propositions. All operations in the experimental
test belong to the Clifford group subset of quantum gates and therefore can be
efficiently simulated classically \cite{Niel2000,Aaro2004,Ande2006}. A
classical device is possible, provided one uses more resources: $N$ classical
bits are required to propagate through the black box in order to specify the
$N$-bit axiomatic set and additional bits are required to model randomness in
measurements corresponding to independent propositions. (Specifically, $2N$
classical bits propagating through the black box are known to be sufficient to
specify definite outcomes in the measurements corresponding to the axioms and
random outcomes in the measurements of fully independent propositions
\cite{Spek2007,Pate2008}.) Such a device can give the truth values of
dependent propositions according to classical logic. On the level of
elementary physical systems, however, the world is known to be quantum. It is
intriguing that nature supplies us with physical systems that can reveal
logical dependence but cannot be used to learn the classical truth values.

A historic point finally deserves comment. The inference that classical logic
cannot capture the structure of quantum mechanics was made by Birkhoff and von
Neumann and started the field of quantum logic \cite{Birk1936}. Our link
between mathematics/logic and certain elements of quantum physics is related
to, but yet distinct from their approach. Quantum logic was invented to
provide an understanding of quantum physics in terms of a set of non-classical
logical rules for propositions which are identified with projective quantum
measurements. However, \textquotedblleft one requires the entire theoretical
machinery of quantum mechanics to justify quantum logic\textquotedblright%
\ \cite{Pito1989}. Our approach aims at providing a justification for quantum
randomness starting from an operational representation of purely mathematical
propositions and systems with limited information content.

The no-go theorems of Bell \cite{Bell1964} and Kochen and Specker
\cite{Koch1967} prove that quantum randomness cannot be understood as stemming
from the ignorance of a hidden variable substructure without coming into
conflict with local realism and non-contextuality. This suggests that quantum
randomness might be of \textit{irreducible} (objective) nature
\cite{Svoz1990,Calu2005b} and a consequence of fundamentally limited
information content of physical systems, namely $N$ bits in $N$
qubits~\cite{Zeil1999}. If one adopts this view, the present work explains in
which experiments the outcomes will be irreducibly random, namely in those
that correspond to logically independent propositions.

After leaving the black boxes the $N$ qubits' quantum states encode exactly
$N$ bits of information about Boolean functions, i.e.\ the systems encode an
$N$-bit\textit{ axiom}, and the other logically complementary propositions are
independent of this axiom. If there exists no underlying (hidden variable)
structure, no information is left for specifying their truth values. However,
the qubits can be measured in the bases corresponding to independent
propositions, and -- as in any measurement -- will inevitably give outcomes,
e.g.\ \textquotedblleft clicks\textquotedblright\ in detectors. These clicks
must not contain any information whatsoever about the truth value of the
independent proposition. Therefore, the individual quantum outcomes must be
random, reconciling logical independence with the fact that a quantum system
always gives an \textquotedblleft answer\textquotedblright\ when
\textquotedblleft asked\textquotedblright\ in an experiment. This provides an
intuitive understanding of quantum randomness, a key quantum feature, using
mathematical reasoning. Moreover, the same argument implies that randomness
necessarily occurs in any physical theory of systems with limited information
content in which measurements are operationally identified with asking
questions about independent propositions \cite{Pate2008b}.

In conclusion, we have demonstrated that the dependence or independence of
certain mathematical propositions in a finite axiomatic set can be tested by
performing corresponding Pauli group measurements. (It would be interesting to
investigate the possiblity of extending our results beyond this class of
measurements.) This is achieved via an isomorphism between axioms and quantum
states as well as between propositions and quantum measurements. Dependence
(Independence) is revealed by definite (random) outcomes. Having this
isomorphism, logical independence needs not to be proved by logic but can be
inferred from experimental results. From the foundational point of view, this
sheds new light on the (mathematical) origin of quantum randomness in these
measurements. Under the assumption that the information content of $N$
elementary physical systems (i.e.\ qubits) is \textit{fundamentally
restricted} to $N$ bits such that no underlying (hidden variable) structure
exists, measurement outcomes corresponding to logically independent
propositions must be irreducibly random.

\textit{Acknowledgements}. We are grateful to G. J. Chaitin for discussions.
We acknowledge financial support from the Austrian Science Fund (FWF), the
Doctoral Program CoQuS (FWF), the European Commission under the Integrated
Project Qubit Applications (QAP) funded by the IST directorate, the Marie
Curie Research Training Network EMALI, the IARPA-funded U.S. Army Research
Office, and the Foundational Questions Institute (FQXi).

\subsection{Appendix}

\begin{figure}[b]
\begin{center}
\includegraphics[width=.4\textwidth]{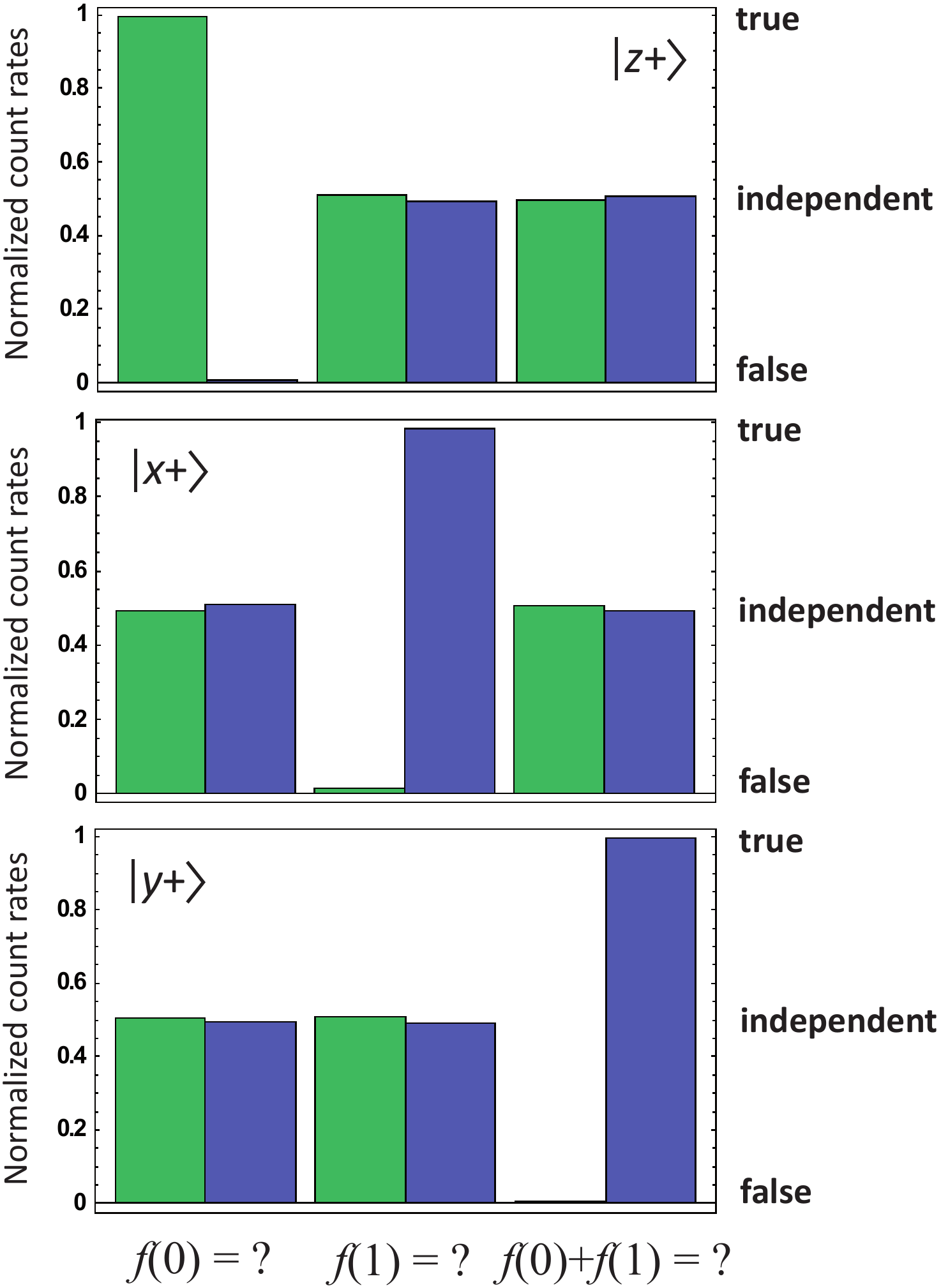}
\end{center}
\par
\vspace{-0.25cm}\caption{We input the qubit in a well-defined Pauli operator
eigenstate $\left\vert z+\right\rangle $, $\left\vert x+\right\rangle $, or
$\left\vert y+\right\rangle $ into the black box, shown from top to bottom.
The black box encodes two classical bits, $f(0)$ and $f(1)$, and the
measurement is chosen such that the single bit $f(0)$, $f(1)$, or $f(0)+f(1)$,
is read out. For every input state we measure in all three complementary
bases, i.e., $z$ [asking for $f(0)$], $x$ [$f(1)$], and $y$ [$f(0)+f(1)$],
shown from left to right. The three measurements are related to three
logically complementary questions (A), (B), (C) of the main text as indicated
by the labels. This particular plot is the experimentally obtained data for
the black box realizing the function $y_{1}$. Similar results were obtained
for the other black box configurations $y_{0}$, $y_{2}$, and $y_{3}$. Green
(blue) bars represent outcomes \textquotedblleft0\textquotedblright%
\ (\textquotedblleft1\textquotedblright) in the respective detectors, giving
the answer to the corresponding question. Each input state, after leaving the
black box, reveals the truth value of one and only one of the propositions,
i.e.\ it encodes a one-bit axiom. Given this axiom, the remaining two
logically complementary propositions are independent. This independence is
revealed by complete randomness of the outcomes in the other two measurement
bases. Statistical errors are at most 0.03\thinspace\% in each graph and
therefore not visible.}%
\label{Figure_Undecidability_Q1}%
\end{figure}Here, we describe experiments which were conducted in order to
illustrate the concepts developed in the main text. In the case of a single
Boolean function $f(x)$, we use the polarization of single photons as
information carriers of binary properties encoded by the configuration in the
\textquotedblleft black box\textquotedblright. The single photons are
generated in the process of spontaneous parametric down-conversion (SPDC)
\cite{Kwia1995}. The horizontal/vertical linear, $+$45$%
{{}^\circ}%
$/$-$45$%
{{}^\circ}%
$ linear, right/left circular polarization of the photon corresponds to
eigenstates $\left\vert z\pm\right\rangle $, $\left\vert x\pm\right\rangle $,
and $\left\vert y\pm\right\rangle $ of the Pauli operators, respectively. We
start by initializing the qubit in a definite polarization state by inserting
a linear polarizer in the beam path. The qubit then propagates through the
black box in which the Boolean functions are encoded with the help of
half-wave plates (HWP) which implement the product of Pauli operators
$\hat{\sigma}_{x}^{f(0)}\,\hat{\sigma}_{z}^{f(1)}$, eq.\ (\ref{eq U}).
Subsequently, measurements of $\hat{\sigma}_{z}$, $\hat{\sigma}_{x}$, and
$\hat{\sigma}_{y}$, which test the truth value of a specific proposition, are
performed as projective measurements in the corresponding polarization basis.
Specifically, to perform $\hat{\sigma}_{z}$ measurements we use a polarizing
beam-splitter (PBS) whose output modes are fiber-coupled to single-photon
detector modules and use wave plates in front of the PBS to change the
measurement basis. The truth value of the proposition now corresponds to
photon detection in one of the two output modes of the PBS.

First, we confirm that complementary quantum measurements indeed reveal truth
values of respective logically complementary propositions. To achieve this we
prepare the system in a state belonging to the basis in which we finally
measure. Specifically, we verify that a measurement in the $z$ basis gives the
value of $f(0)$ and similarly, measurements in $x$ and $y$ bases give the
value of $f(1)$ and $f(0)+f(1)$, respectively.

Next, we demonstrate that dependent (independent) propositions are identified
by a sequence of definite (random) outcomes of quantum measurements. For each
of the three choices of the initial state, we \textquotedblleft
ask\textquotedblright\ all three logically complementary questions by
measuring in all three different complementary bases.
Figure~\ref{Figure_Undecidability_Q1} shows that for every input state one and
only one question has a definite answer. This is the axiom encoded in the
system leaving black box. The remaining propositions are independent given
that axiom. This is signified by the observation that the corresponding
measurement outcomes are completely random, i.e.\ evenly distributed.

\begin{figure}[t]
\begin{center}
\includegraphics[width=.6\textwidth]{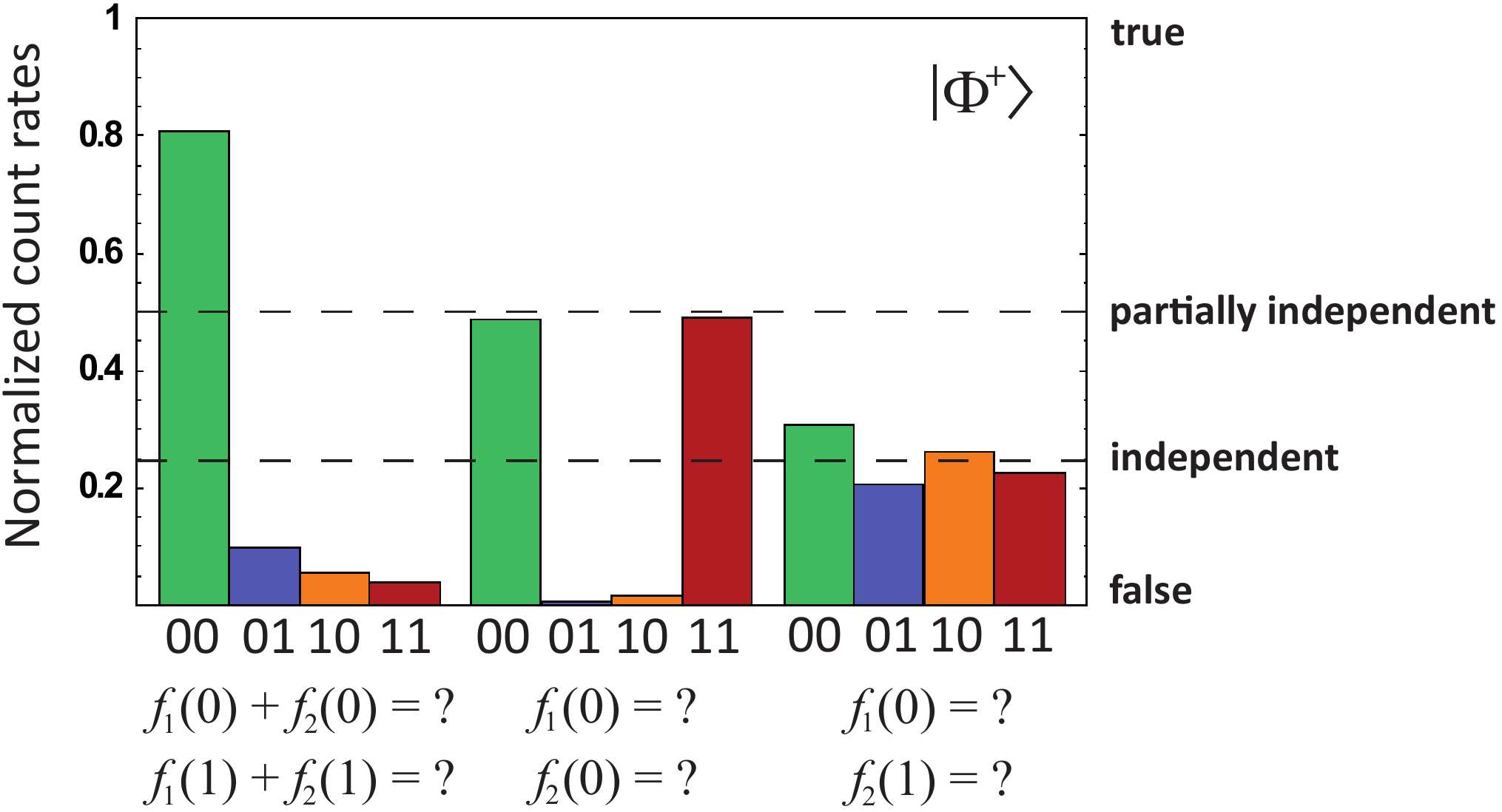}
\end{center}
\par
\vspace{-0.25cm}\caption{In this two-qubit experiment a $\left\vert \Phi
^{+}\right\rangle $ Bell state is measured in three different bases (the Bell
basis as well as $\left\vert z\pm\right\rangle _{1}\left\vert z\pm
\right\rangle _{2}$ and $\left\vert z\pm\right\rangle _{1}\left\vert
x\pm\right\rangle _{2}$, shown from left to right). Plotted are the normalized
count rates associated with the relevant detector combinations for the black
box encoding function $y_{2}$ on both photons. The first (second) label gives
the answer to the upper (lower) question. The results of the measurements
performed in the Bell basis show that the two qubits encode proposition (E) of
the main text. The data in the middle plot reveal partial independence of
proposition (F), given (E) as an axiom, as indicated by the random outcomes in
two out of four detector combinations. In contrast, the right plot presents
the data corresponding to the fully independent proposition (D), where the
outcomes are completely random. Similar results for propositions (E), (F) and
(D) were obtained for other black box encodings. The reason for the fact that
the Bell state in the left plot is not identified with unit fidelity stems
from imperfections in the experimental setup (cf.\ Reference \cite{Walt2005b}%
). Unequal detector efficiencies explain the small bias in the right plot.
Statistical errors are at most 2\thinspace\% for the left plot and at most
0.1\thinspace\% for the other plots and therefore not visible.}%
\label{Figure_Undecidability_Q2}%
\end{figure}In the case of two Boolean functions, $f_{1}(x)$ and $f_{2}(x)$,
we prepare two-photon states with the help of SPDC. The encoding of these
functions within the black boxes is done akin to the single-qubit case. We
start our investigation by confirming that the truth values of the
(elementary) propositions in (E), (F) and (D) are revealed by measurements
performed in the bases $b_{\text{E}}$, $b_{\text{F}}$ and $b_{\text{D}}$ of
the main text, respectively. As in the single-qubit case, preparation and
measurement are in the same basis (results not shown). Next, we prepare the
$\left\vert \Phi^{+}\right\rangle $ Bell state and measure it in the bases
$b_{\text{E}}$, $b_{\text{F}}$ and $b_{\text{D}}$. As can be seen in the left
plot of Figure~\ref{Figure_Undecidability_Q2}, measurements in the Bell basis,
$b_{\text{E}}$, prove that the entangled state indeed encodes joint properties
of the functions $f_{1}(x)$ and $f_{2}(x)$, i.e.\ information about (E). These
joint two-qubit measurements require a so-called Bell State Analyzer (BSA)
\cite{Wein1994,Mich1996,Walt2005b}, the heart of which is a non-linear gate,
such as a controlled-NOT gate \cite{Bare1995,Pitt2002,OBri2003,Gasp2004}. For
experimental details see Reference~\cite{Walt2005b}. Measurements in other
bases can then be interpreted in terms of \textquotedblleft
partial\textquotedblright\ and \textquotedblleft full\textquotedblright%
\ independence. Proposition (D) is fully independent given (E) as an axiom
encoded into the photons leaving the black boxes. This can be seen from the
right part in Figure~\ref{Figure_Undecidability_Q2}, in which all four
measurement outcomes occur with equal probability. On the other hand,
proposition (F) is partially independent. This is experimentally revealed by
the count distribution of the middle part in
Figure~\ref{Figure_Undecidability_Q2}. The partial independence is disclosed
by the randomness of the two occurring outcomes, while the other two outcomes
do not appear.


\begin{thebibliography}{99}                                                                                               %


\bibitem {Koch1967}S. Kochen and E. Specker, Journal of Mathematics and
Mechanics \textbf{17}, 59 (1967).

\bibitem {Bell1964}J. S. Bell, Physics (N.Y.) \textbf{1}, 195 (1964).

\bibitem {Chai1982}G. J. Chaitin, Int. J. Theor. Phys. \textbf{21}, 941 (1982).

\bibitem {Calu2005}C. S. Calude and H. Jürgensen, Adv. Appl. Math.
\textbf{35}, 1 (2005).

\bibitem {Hole1973}A. S. Holevo, Probl. Inf. Transm. \textbf{9}, 177 (1973).

\bibitem {Zeil1999}A. Zeilinger, Found. Phys. \textbf{29}, 631 (1999).

\bibitem {Goed1931}K. Gödel, Monatsheft für Mathematik und Physik (Akademische
Verlagsgesellschaft Leipzig) \textbf{38}, 173 (1931).

\bibitem {Nage1960}E. Nagel and J. R. Newman, \textit{Gödel's proof} (New York
University Press, 1960).

\bibitem {Kwia1995}P. G. Kwiat, K. Mattle, H. Weinfurter, A. Zeilinger, A. V.
Sergienko, and Y. Shih, Phys. Rev. Lett. \textbf{75}, 4337 (1995).

\bibitem {Gott1996}D. Gottesman, Phys. Rev. A \textbf{54}, 1862 (1996).

\bibitem {Niel2000}M. A. Nielsen and I. L. Chuang, \textit{Quantum Computation
and Quantum Information} (Cambridge University Press, 2000).

\bibitem {Raus2003}R. Raussendorf, D. E. Browne, and H. J. Briegel, Phys. Rev.
A \textbf{68}, 022312 (2003).

\bibitem {Wein1994}H. Weinfurter, Europhys. Lett. \textbf{25}, 559 (1994).

\bibitem {Gree1989}D. Greenberger, M. A. Horne, and A. Zeilinger,
in:~\textit{Bell's Theorem, Quantum Theory, and Conceptions of the Universe},
ed.\ M. Kafatos (Kluwer Academic Publishers, 1989); electronic version:
arXiv:0712.0921v1 [quant-ph].

\bibitem {Merm1990}N. D. Mermin, Phys. Rev. Lett. \textbf{65}, 1838 (1990).

\bibitem {Pan2000}J.-W. Pan, D. Bouwmeester, M. Daniell, H. Weinfurter, and A.
Zeilinger, Nature \textbf{403}, 515 (2000).

\bibitem {Pere1995}A. Peres, \textit{Quantum Theory:\ Concepts and Methods}
(Kluwer Academic Publishers, 1995).

\bibitem {Aaro2004}S. Aaronson and D. Gottesman, Phys. Rev. A \textbf{70},
052328 (2004).

\bibitem {Ande2006}S. Anders and H. J. Briegel, Phys. Rev. A \textbf{73},
022334 (2006).

\bibitem {Spek2007}R. Spekkens, Phys. Rev. A \textbf{75}, 032110 (2007).

\bibitem {Pate2008}T. Paterek, B. Daki\'{c}, and \v{C}. Brukner, Phys. Rev. A
\textbf{79}, 012109 (2009).

\bibitem {Birk1936}G. Birkhoff and J. von Neumann, Ann. Math. \textbf{37}, 823 (1936).

\bibitem {Pito1989}I. Pitowski, \textit{Quantum Probability -- Quantum Logic}
(Springer, 1989).

\bibitem {Svoz1990}K. Svozil, Phys. Lett. A \textbf{143}, 433 (1990).

\bibitem {Calu2005b}C. S. Calude and M. A. Stay, Int J. Theor. Phys.
\textbf{44}, 1053 (2005).

\bibitem {Pate2008b}T. Paterek, B. Daki\'{c}, and \v{C}. Brukner,
arXiv:0804.1423 [quant-ph].

\bibitem {Mich1996}M. Michler, K. Mattle, H. Weinfurter, and A. Zeilinger,
Phys. Rev. A \textbf{72}, 1209(R) (1996).

\bibitem {Walt2005b}P. Walther and A. Zeilinger, Phys. Rev. A \textbf{72},
010302(R) (2005).

\bibitem {Bare1995}A. Barenco, C. H. Bennett, R. Cleve, D. P. DiVincenzo, N.
Margolus, P. Shor, T. Sleator, J. Smolin, and H. Weinfurter, Phys. Rev. A
\textbf{52}, 3457 (1995).

\bibitem {Pitt2002}T. B. Pittman, B. C. Jacobs, and J. D. Franson, Phys. Rev.
Lett. \textbf{88}, 257902 (2002).

\bibitem {OBri2003}J. L. O'Brien, G. J. Pryde, A. G. White, T. C. Ralph, and
D. Branning, Nature \textbf{426}, 264 (2003).

\bibitem {Gasp2004}S. Gasparoni, J.-W. Pan, P. Walther, T. Rudolph, and A.
Zeilinger, Phys. Rev. Lett. \textbf{92}, 020504 (2004).
\end{thebibliography}
\end{document}